\documentclass[letterpaper,twocolumn,english,prc,aps,showpacs,superscriptaddress]{revtex4}

\usepackage{ae,aecompl}
\usepackage[T1]{fontenc}
\usepackage[latin9]{inputenc}
\usepackage{babel}
\usepackage{units}
\usepackage{amsmath}
\usepackage{amssymb}
\usepackage{graphicx}
\usepackage{subfigure}
\usepackage{tikz}
\usepackage{multirow}
\usepackage{ulem}

\makeatletter

\DeclareTextFontCommand{\helvetica}{\fontfamily{phv}\selectfont}

\makeatother

\begin{document}


\title{A new ``3D Calorimetry'' of hot nuclei }

\author{E. Vient}
\thanks{vient@lpccaen.in2p3.fr; http://caeinfo.in2p3.fr/}
\affiliation{Normandie Univ, ENSICAEN, UNICAEN, CNRS/IN2P3, LPC Caen, F-14000 Caen, France}

\author{L. Manduci}
\affiliation{\'Ecole des Applications Militaires de l'\'Energie Atomique, B.P. 19, F-50115 Cherbourg, France}
\affiliation{Normandie Univ, ENSICAEN, UNICAEN, CNRS/IN2P3, LPC Caen, F-14000 Caen, France}

\author{E. Legou\'ee}
\affiliation{Normandie Univ, ENSICAEN, UNICAEN, CNRS/IN2P3, LPC Caen, F-14000 Caen, France}

\author{L. Augey}
\affiliation{Normandie Univ, ENSICAEN, UNICAEN, CNRS/IN2P3, LPC Caen, F-14000 Caen, France}

\author{E. Bonnet}
\affiliation{SUBATECH UMR 6457, IMT Atlantique, Universit\'e de Nantes, CNRS-IN2P3, 44300 Nantes, France}

\author{B. Borderie}
\affiliation{Institut de Physique Nucl\'eaire, CNRS/IN2P3, Univ. Paris-Sud, Universit\'e Paris-Saclay, F-91406 Orsay cedex, France}

\author{R. Bougault}
\affiliation{Normandie Univ, ENSICAEN, UNICAEN, CNRS/IN2P3, LPC Caen, F-14000 Caen, France}

\author{A. Chbihi}
\affiliation{Grand Acc\'el\'erateur National d'Ions Lourds (GANIL), CEA/DRF-CNRS/IN2P3, Bvd. Henri Becquerel, 14076 Caen, France}

\author{D. Dell'Aquila}
\affiliation{Institut de Physique Nucl\'eaire, CNRS/IN2P3, Univ. Paris-Sud, Universit\'e Paris-Saclay, F-91406 Orsay cedex, France}
\affiliation{Dipartimento di Fisica 'E. Pancini' and Sezione INFN, Universit\'a di Napoli 'Federico II', I-80126 Napoli, Italy}

\author{Q. Fable}
\affiliation{Grand Acc\'el\'erateur National d'Ions Lourds (GANIL), CEA/DRF-CNRS/IN2P3, Bvd. Henri Becquerel, 14076 Caen, France}

\author{L. Francalanza}
\affiliation{Dipartimento di Fisica 'E. Pancini' and Sezione INFN, Universit\'a di Napoli 'Federico II', I-80126 Napoli, Italy}

\author{J.D. Frankland}
\affiliation{Grand Acc\'el\'erateur National d'Ions Lourds (GANIL), CEA/DRF-CNRS/IN2P3, Bvd. Henri Becquerel, 14076 Caen, France}

\author{E. Galichet}
\affiliation{Institut de Physique Nucl\'eaire, CNRS/IN2P3, Univ. Paris-Sud, Universit\'e Paris-Saclay, F-91406 Orsay cedex, France}
\affiliation{Conservatoire National des Arts et M\'etiers, F-75141 Paris Cedex 03, France}

\author{D. Gruyer}
\affiliation{Normandie Univ, ENSICAEN, UNICAEN, CNRS/IN2P3, LPC Caen, F-14000 Caen, France}
\affiliation{Sezione INFN di Firenze, Via G. Sansone 1, I-50019 Sesto Fiorentino, Italy}

\author{D. Guinet}
\affiliation{IPNL/IN2P3 et Universit\'e de Lyon/Universit\'e Claude Bernard Lyon1, 43 Bd du 11 novembre 1918 F69622 Villeurbanne Cedex, France}

\author{M. Henri}
\affiliation{Normandie Univ, ENSICAEN, UNICAEN, CNRS/IN2P3, LPC Caen, F-14000 Caen, France}

\author{M. La Commara}
\affiliation{Dipartimento di Fisica 'E. Pancini' and Sezione INFN, Universit\'a di Napoli 'Federico II', I-80126 Napoli, Italy}

\author{G. Lehaut}
\affiliation{Normandie Univ, ENSICAEN, UNICAEN, CNRS/IN2P3, LPC Caen, F-14000 Caen, France}

\author{N. Le Neindre}
\affiliation{Normandie Univ, ENSICAEN, UNICAEN, CNRS/IN2P3, LPC Caen, F-14000 Caen, France}

\author{I. Lombardo}
\affiliation{Dipartimento di Fisica 'E. Pancini' and Sezione INFN, Universit\'a di Napoli 'Federico II', I-80126 Napoli, Italy}
\affiliation{INFN - Sezione Catania, via Santa Sofia 64, 95123 Catania, Italy}

\author{O. Lopez}
\affiliation{Normandie Univ, ENSICAEN, UNICAEN, CNRS/IN2P3, LPC Caen, F-14000 Caen, France}

\author{P. Marini}
\affiliation{CEA, DAM, DIF, F-91297 Arpajon, France}

\author{M. P\^arlog}
\affiliation{Normandie Univ, ENSICAEN, UNICAEN, CNRS/IN2P3, LPC Caen, F-14000 Caen, France}
\affiliation{Hulubei National Institute for R$\And$D in Physics and Nuclear Engineering (IFIN-HH), P.O.BOX MG-6, RO-76900 Bucharest-M\`agurele, Romania}

\author{M. F. Rivet}
\thanks{deceased}
\affiliation{Institut de Physique Nucl\'eaire, CNRS/IN2P3, Univ. Paris-Sud, Universit\'e Paris-Saclay, F-91406 Orsay cedex, France}

\author{E. Rosato}
\thanks{deceased}
\affiliation{Dipartimento di Fisica 'E. Pancini' and Sezione INFN, Universit\'a di Napoli 'Federico II', I-80126 Napoli, Italy}
\author{R. Roy}
\affiliation{Laboratoire de Physique Nucl\'eaire, Universit\'e Laval, Qu\'ebec, Canada G1K 7P4}

\author{P. St-Onge}
\affiliation{Laboratoire de Physique Nucl\'eaire, Universit\'e Laval, Qu\'ebec, Canada G1K 7P4}
\affiliation{Grand Acc\'el\'erateur National d'Ions Lourds (GANIL), CEA/DRF-CNRS/IN2P3, Bvd. Henri Becquerel, 14076 Caen, France}

\author{G. Spadaccini}
\affiliation{Dipartimento di Fisica 'E. Pancini' and Sezione INFN, Universit\'a di Napoli 'Federico II', I-80126 Napoli, Italy}

\author{G. Verde}
\affiliation{Institut de Physique Nucl\'eaire, CNRS/IN2P3, Univ. Paris-Sud, Universit\'e Paris-Saclay, F-91406 Orsay cedex, France}
\affiliation{INFN - Sezione Catania, via Santa Sofia 64, 95123 Catania, Italy}

\author{M. Vigilante}
\affiliation{Dipartimento di Fisica 'E. Pancini' and Sezione INFN, Universit\'a di Napoli 'Federico II', I-80126 Napoli, Italy}

\date{\today}%

\begin{abstract}
In the domain of Fermi energy, it is extremely complex to isolate experimentally fragments and particles issued from the cooling of a hot nucleus produced during a heavy ion collision. This paper presents a new method to characterize more precisely hot Quasi-Projectiles. It tries to take into account as accurately as possible the distortions generated by all the other potential participants in the nuclear reaction. It is quantitatively shown that this method is a major improvement respect to classic calorimetries used with a 4$\pi$ detector array. By detailing and  deconvolving the different steps of the reconstitution of the hot nucleus, this study shows also the respective role played by the experimental device and the event selection criteria on the quality of the determination of QP characteristics.
\end{abstract}

\maketitle
\section{\label{sec1}INTRODUCTION}
The only way to study experimental nuclear thermodynamics is to produce hot nuclei during nuclear collisions. The hot nuclei are obtained thus in extremely violent and complex conditions. In the domain of Fermi energy, we observe clearly for the energy dissipation a competition between nuclear mean field and nucleon-nucleon interaction \cite{Lehaut1, Lopez1}.  The collisions present mostly a strong binary character preserving a very strong memory of the entrance channel \cite{Metivier1}. The process of deeply inelastic diffusion becomes the dominant phenomenon \cite{Steck1, Casini1, Charity1, Jouan1, Gingras1}. It is accompanied by an important emission of light particles, thermally unbalanced \cite{dore1, Gingras1, Peter1}. But we also observe an important production of \textbf{I}ntermediate \textbf{M}ass \textbf{F}ragments (\textbf{IMFs}) at the interface of the two colliding nuclei. This latter is usually called \textbf{Neck Emission} \cite{DiToro1,Gingras1, Stuttge1, Casini2, Montoya1, Lecolley1}. Fusion is also observed but its cross section becomes low for symmetric collisions above 30 A.MeV \cite{Eudes2,Metivier1, Jeong1, Marie1, Beaul1}. Obviously, the respective cross sections of these various processes change according to the system, the incident energy  and the impact parameter. The formed hot nuclei de-excite in many particles and fragments. Only a $4\pi$ array can allow to study such physical processes because of its detection capabilities. We know that it is  fundamental in thermodynamics or statistical mechanics that the studied system has to be perfectly defined and characterized. This is the major experimental challenge encountered by nuclear physicists working in the domain of Fermi energy. For this reason, the main purpose of this paper is to present, understand and validate a new method of characterization of an excited \textbf{Q}uasi-\textbf{P}rojectile (\textbf{QP}) with a $4\pi$ experimental set-up in this energy range. By characterization, we mean: to determine its charge, mass, velocity  and  excitation energy. We want also to improve, possibly,  these ``measurements'' compared to the existing methods \cite{Viola2}. This study is made with the $4\pi$ array INDRA \cite{Pouthas1} for the system Xe + Sn at 50 A.MeV. In the first section, we will discuss how to select the products of the Quasi-Projectile decay among all the particles produced during the reaction. In the following section, we will present the principles of our calorimetry. In the last section, we will study this new experimental calorimetry by comparing to a standard calorimetry. 

\section{IMPROVED SELECTION OF THE EVAPORATED PARTICLES  \label{sec2}}
\subsection{ Light particle characterization : influences of the reference frame and of the experimental set-up \label{ssec2.1}}
Here we discuss firstly how the reference frame used in the analysis and the experimental set-up may influence the spatial and energetic characteristics of the \textbf{L}ight \textbf{C}harged \textbf{P}articles (\textbf{LCPs}) and consequently the determination of the excitation energy of the hot nuclei and the energetic spectrum slopes of these particles.   
\begin{figure} [htbp]
\centerline {\includegraphics[width=8.7cm, height=3.64cm] {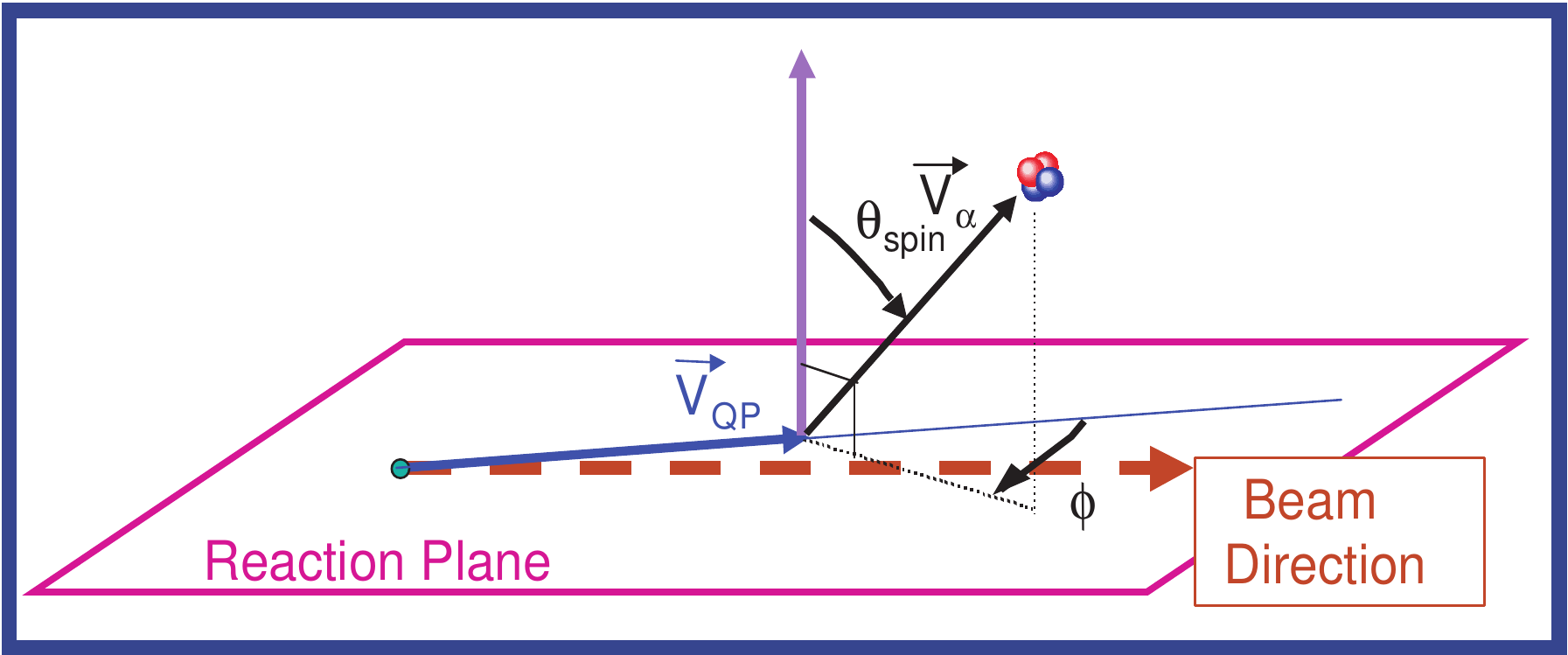}}
\caption {Definitions of the polar angle $\theta_{spin}$ and of the azimuthal angle $\phi$ of light emitted particle  (the azimuthal angle $\phi$  presented in the figure is negative).\label{fig1}}
\end{figure}
For this study, the SIMON event generator, developed by D.Durand \cite{Durand1}, was used. It is set to supply only pure binary collisions Xe + Sn at 50 A.MeV without pre-equilibrium particles.
We will study different angular distributions of emitted particles. Figure \ref{fig1} shows the frame used to study the LCP angular distributions and allows to define the different used angles.
The polar angle $\theta_ {spin}$ is defined as the angle between the vector normal to the reaction plane and the velocity vector of a particle emitted by the Quasi-Projectile in the QP frame. The azimuthal angle $\phi$ is defined as the angle between the QP velocity vector in the frame of the center of mass (c.m.) and the normal projection of the velocity vector of the emitted particle on the reaction plane.
The azimuthal angle is defined positive when the projection of the velocity vector of the particle on the reaction plane is located on the ``left'' with respect to the direction of the QP velocity vector. A particle emitted in the reaction plane and in the direction of the QP velocity vector  will have $\phi=0^{\circ}$ and $\theta_{spin}=90^{\circ}$.
\begin{figure} [ht]
\centerline {\includegraphics[width=9.2cm, height=15.9cm] {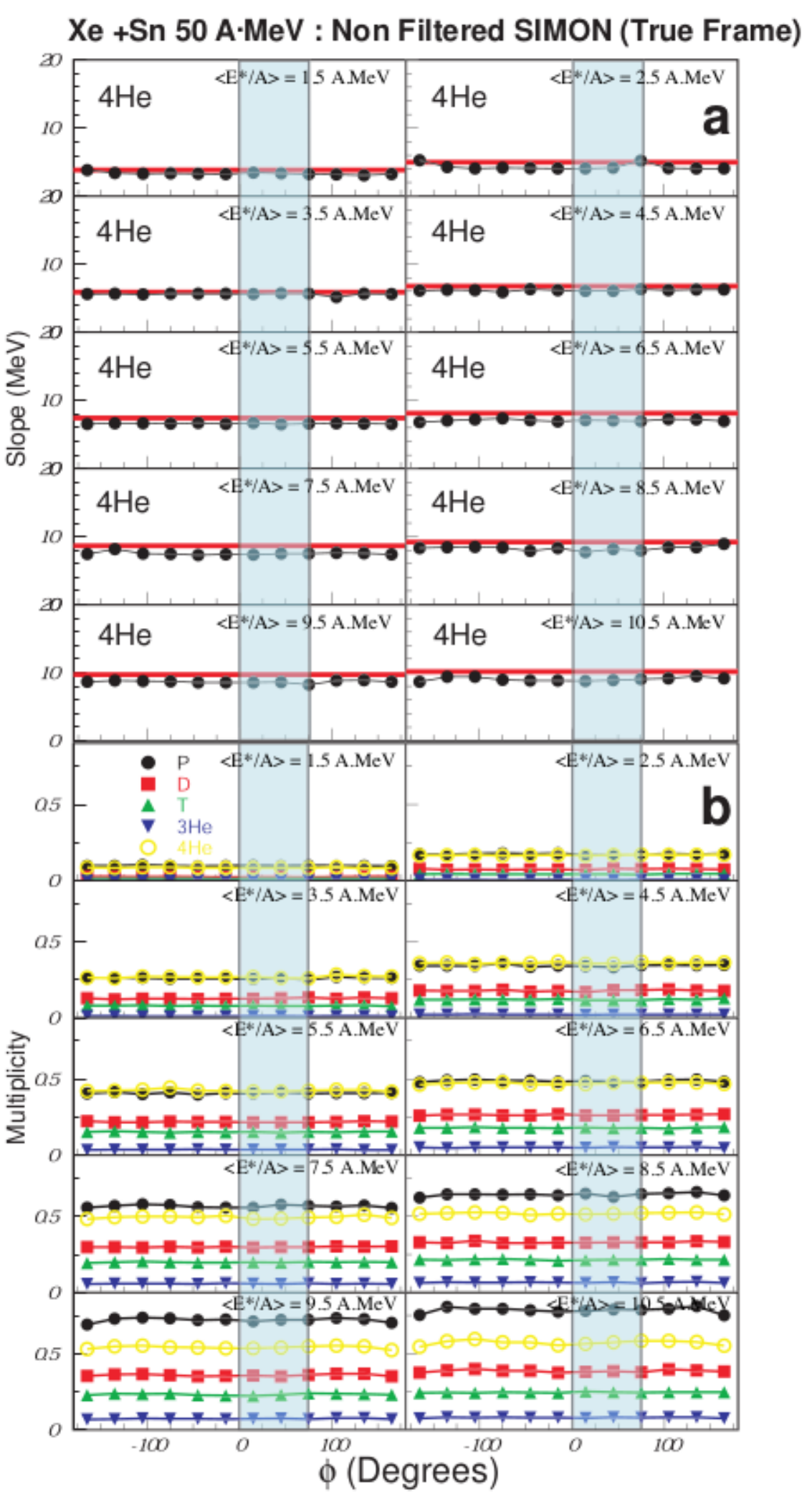}}
\caption {\textbf{a)} Slopes of energy spectra of alphas evaporated by the QP, obtained by fit for various domains of $\phi$, according to the associated mean $\phi$. This is done for various selections according to $E^{*}/A$ of the QP. The red line corresponds to the true initial temperature of the QP, associated with each range of excitation energy per nucleon.\textbf{ b)} Mean multiplicities of the light particles emitted by the QP, generated by SIMON, according to $\phi$, without experimental filter (the velocity vector being calculated in the true initial frame of the QP). \label{fig2}}
\end{figure}
Figure \ref{fig2} shows two graphs built in the real initial frame of the emitting source, \textit{i.e.} the QP  (which we will call ``True Frame''). The events are not filtered at this stage. Therefore, we assume the use of a perfect detector. The particles were solely evaporated by the QP according to SIMON. The  graphs  of the figure \ref{fig2}-a show the slopes obtained by a maxwellian fit of the energy spectra of the alphas, for various angular domains of $\phi$ and for ten bins of 1 MeV of the QP excitation energy per nucleon. The collisions are increasingly violent by going from left to right and downwards in the figure. 
The graph of the figure \ref{fig2}-b presents the mean multiplicity of the light charged particles according to the azimuthal angle $\phi$ for various ranges of the excitation energy per nucleon. 
We find in both cases the expected result for the emission by a thermalized nucleus, \textit{i.e.} flat angular distributions for all the light particles and an uniform measured slope whatever $\phi$.
This guarantees the validity and the coherence of SIMON with respect to the treatment of the isotropic QP decay.
\begin{figure} [htbp]
\centerline {\includegraphics[width=8.6cm, height=15.3cm] {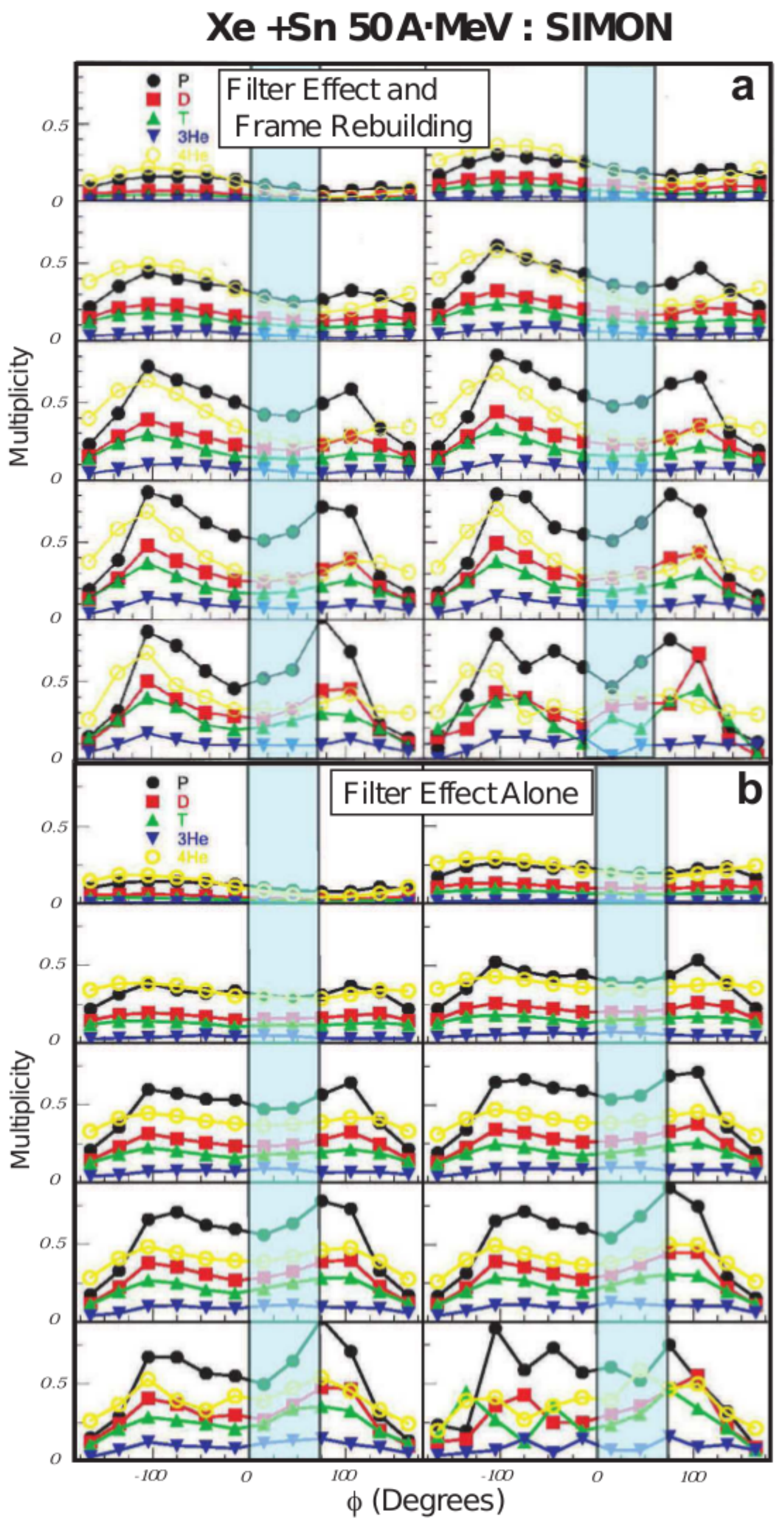}}
\caption {\textbf{a)} $\phi$ distributions of the light particles emitted by the QP supplied by SIMON after the INDRA filter (the velocity vector of the particle being calculated here in the rebuilt frame for various selections according to $E_{tr12}$). \textbf{b)} $\phi$ distributions of the light particles emitted by the QP, generated by SIMON, after passage in the INDRA filter, for various selections according to $E_{tr12}$. The considered events are called ``complete events'' (the velocity vector of each particle being calculated in the true initial frame of the QP). \label{fig3}}
\end{figure}

To see and understand the impact of the used experimental set-up, INDRA,  we coupled to the generated events by SIMON a software which integrates and simulates all the phases of the ion detection by INDRA (detection efficiency, kinetic energy resolution, angular resolution, isotopic identification). This fundamental stage is summarized in terms of the ``INDRA experimental filter''. 
A correct calorimetry requires the complete detection of the nuclear reaction products. To try to achieve this goal, we required a constraint on the total detected charge and momentum (80 \% of the initial value) as done  in \cite{Steck2, Vient1}.  We also aim at understanding the influence of this event selection concerning the calorimetry.  These events are called hereafter  ``complete events''. Indeed, because we do not have an experimental direct access to the excitation energy of the nucleus, we have chosen the total transverse kinetic energy of light charged particles $E_ {tr12}$ as experimental selector of the violence of the collision \cite{Peter1,  Lukasik1}. In the figure \ref{fig3}-b, for the complete events, we can observe precisely the sole influence of the INDRA experimental filter on the spatial distribution of particles evaporated by the QP (the frame used here is the true frame of the QP). This selection of complete events involves an important deformation of the $\phi$ distribution of evaporated particles, even if the velocity vector of evaporated particles is defined in the true initial frame. There are apparently more light particles emitted on the right than on the left with respect to the QP velocity vector, for the peripheral collisions. In fact, the criterion of completeness implies that we only keep events for which the emission of the light particles allowed the QP residue to avoid the forward of the detector, made to let pass the beam.  This effect is all the more strong as the QP is forward focused.

\begin{figure} [htbp]
\centerline{\includegraphics[width=8.348cm, height=9.34cm] {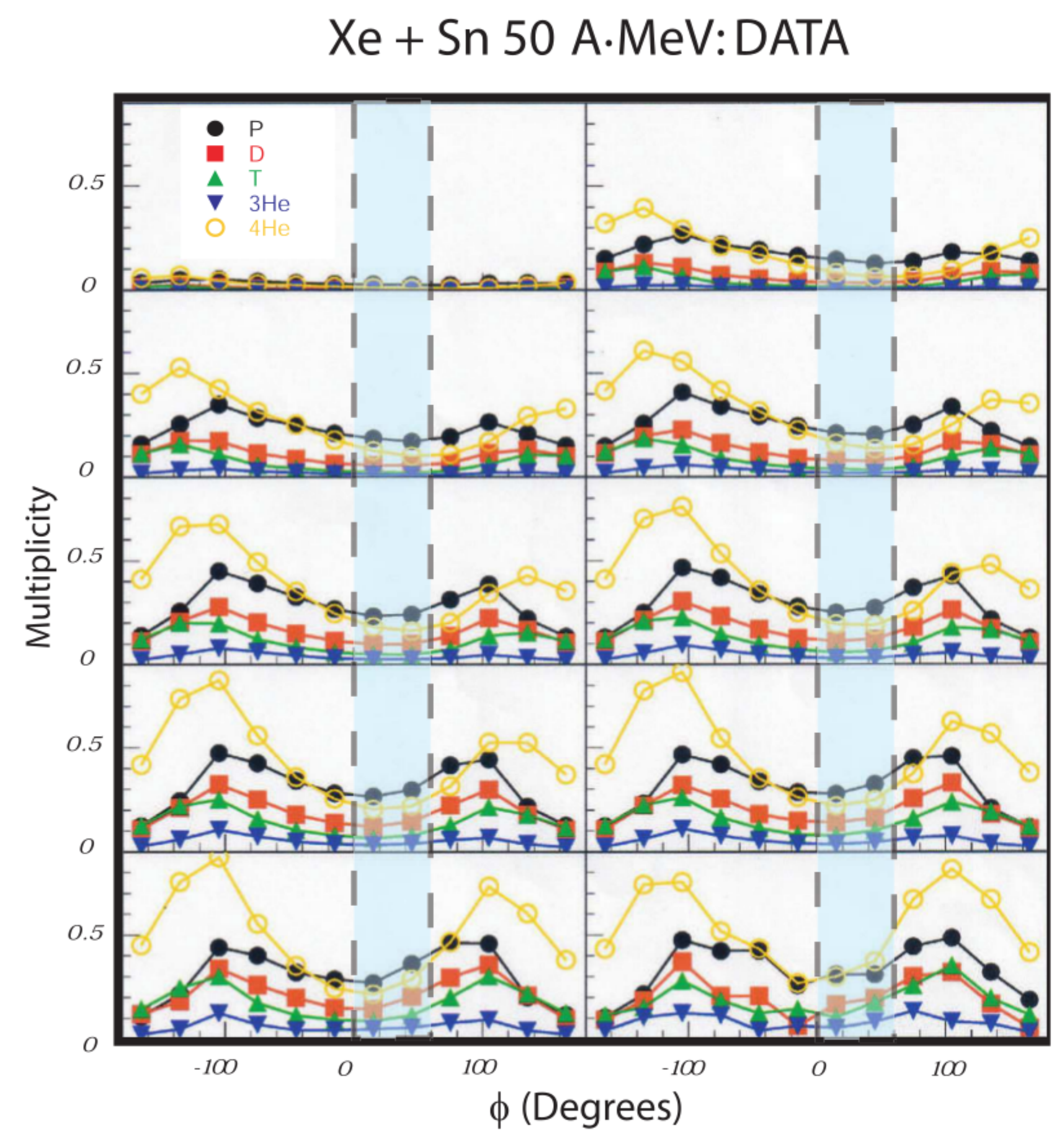}}
\caption{Experimental azimuthal distributions of the light charged particles located in the forward hemisphere of the center of mass (their velocity vector being calculated in the rebuilt frame) for various $E_ {tr12}$ bins, obtained experimentally with INDRA. The considered events are events said ``complete''.\label{fig4}}
\end{figure}
Now, as shown in the figure \ref{fig3}-a,  if the velocity vector is defined in the rebuilt frame (built considering intermediate mass fragments and heavy fragments located in the forward hemisphere respect to the c.m.), we can observe that the effect is even more amplified. What is observed in the figure \ref{fig3} is the ``right-left effect'' described in \cite{Steck2, Vient1}.
This important deformation involves a breaking of the revolution symmetry for the evaporation around the axis passing by the center of the QP and perpendicular to the reaction plan, in the velocity space.

\begin{figure} [htbp]
\centerline{\includegraphics[width=8.9cm, height=16.188cm] {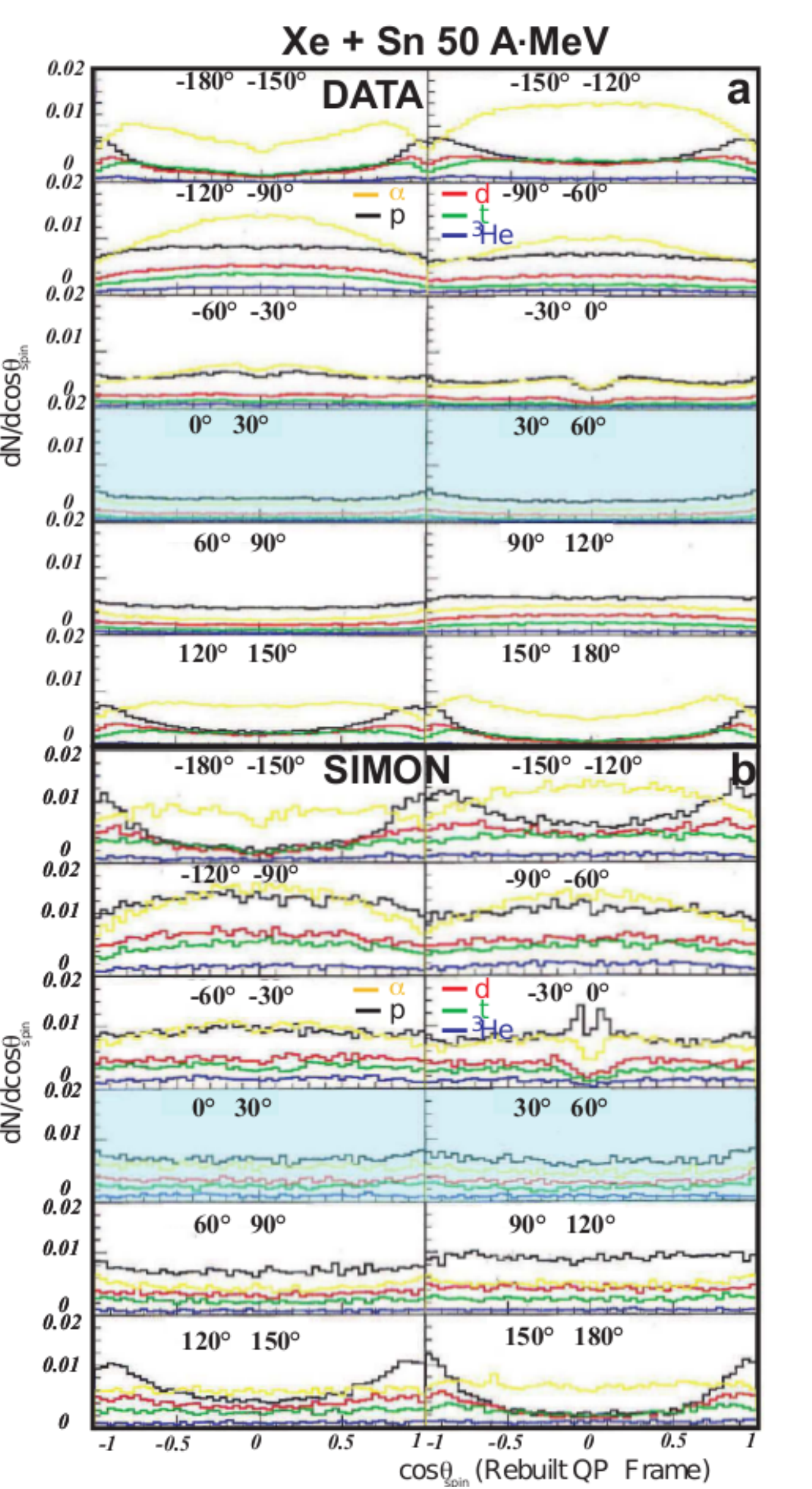}}
\caption{\textbf{a)} Multiplicity distributions of the LCPs, located in the front
 of the center of mass, according to the $cos(\theta_{spin})$ for various  $\phi$ bins. They are obtained for data selection according to the transverse energy corresponding to semi-peripheral collisions and also for complete events.\textbf{ b)} The same distributions for data supplied by SIMON with the same event selection (the frame of the source is reconstructed by the experimental method).\label{fig5}}
\end{figure}

We can wonder whether this apparent effect is just an artifact related to SIMON. This is why, we present in the figure \ref{fig4} the mean multiplicities of the light charged particles, located in the front of the center of mass, according to $ \phi$, obtained with the experimental data recorded by the INDRA array. 
We observe indeed the same trend, even enhanced because of pre-equilibrium emission (not taken into account in SIMON) which is preferentially located between the two partners of the collision, therefore at angles $\phi$ near 180$^{\circ}$.
\begin{figure} [htbp]
\centerline{\includegraphics [width=8.48cm, height=15.28cm] {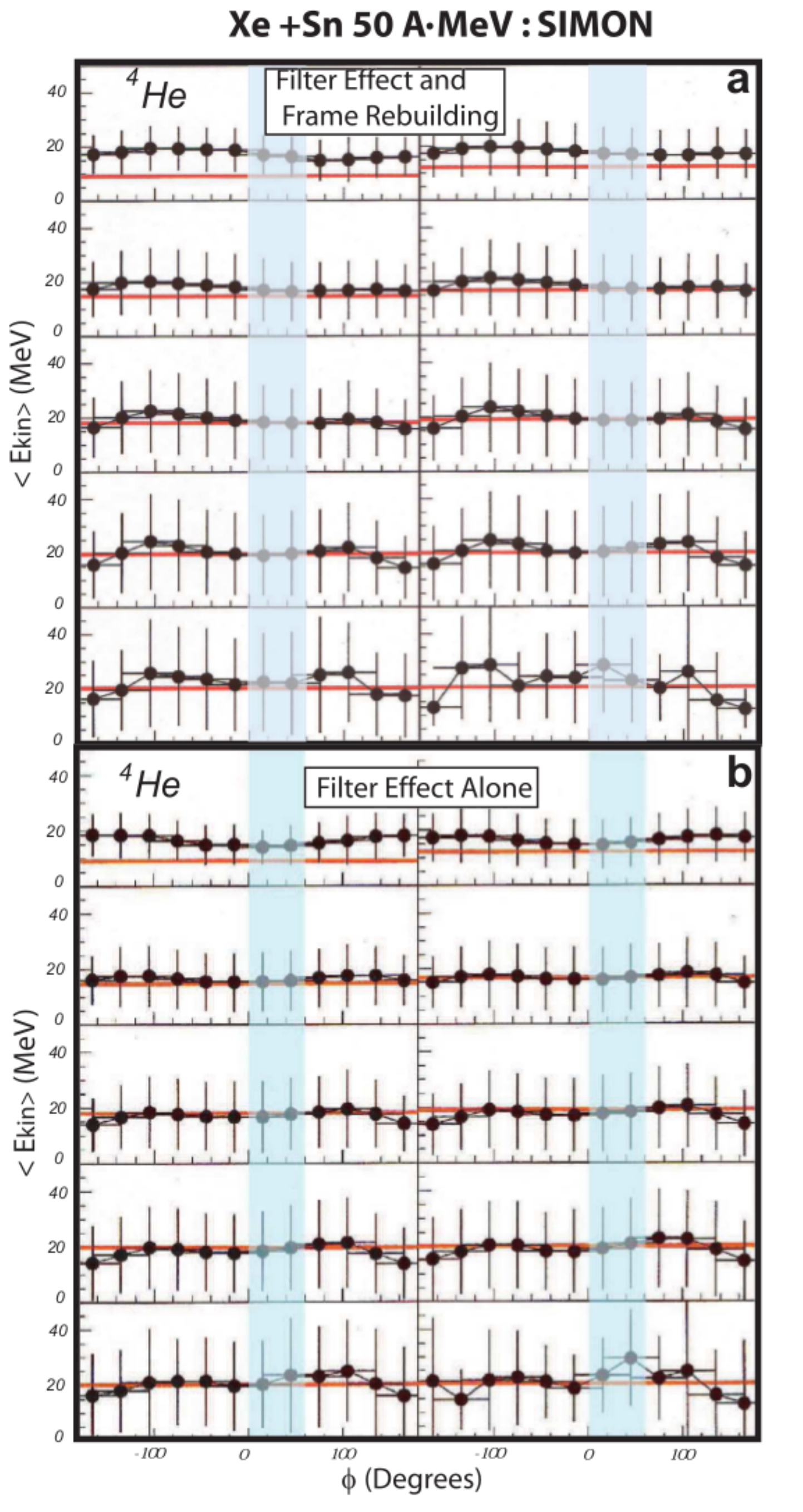}}
\caption{\textbf{a)} Mean energy distributions of the alphas evaporated by the QP according to $\phi$ generated by SIMON after INDRA filter (their velocity vector being calculated in the rebuilt frame) for various $E_{tr12}$ bins. \textbf{b)} Mean energy distributions of the alphas evaporated by the QP according to $\phi$, generated by SIMON after INDRA filter (their velocity vector being calculated in the true initial frame) for various $E_{tr12}$ bins. For \textbf{a) }and \textbf{b)}, the considered events are known as ``complete''. The red line corresponds to a value twice the expected temperature of the hot nucleus (with a density parameter equal to 10).\label{fig6}}
\end{figure} 
 The $\theta_ {spin} $ distributions are also modified by the selection criteria and the INDRA experimental filter. In figure \ref{fig5}, we present a comparison of the multiplicity distributions of particles located in the front of the center of mass according to the $cos(\theta _{spin})$ obtained with data (figure \ref{fig5}-a) and pure binary SIMON simulations (figure \ref{fig5}-b)  for complete events,  we selected semi-peripheral collisions by using the LCP total transverse kinetic energy. We find in both cases similar results : there is no apparent symmetry in $\phi$. Only the angular domains 0$^{\circ}$ - 30$^{\circ}$ and 30$^{\circ}$ - 60$^{\circ}$ are similar.
 In fact, these trends were already seen in a study made by Steckmeyer et al. in \cite{Steck2}, where the authors showed how the ``right-left effect'' and the experimental method of reconstruction (only from the intermediate mass fragments and heavy fragments) deform the symmetric distribution expected for LCP  emissions by an hot rotating source. The required completeness favors the events presenting a pronounced ``right-left effect'', for the peripheral collisions. It was also shown in this study \cite{Steck2} that the area of space, located at the right front in the frame of the reconstructed QP, is strongly polluted by a pre-equilibrium emission (for semi-peripheral and central collisions). There is also a slight pollution of the front of center of mass by the Quasi-Target emission, as already noted in the reference \cite{Vient1}.

Now, for complete events, we will make an equivalent study concerning the energy characteristics of the LCPs in the frame of the emitting nucleus. We study first the effect of the filter (in the figure \ref{fig6}-b) then the cumulated effect of the filter and of the reconstruction of the frame velocity vector (in the figure \ref{fig6}-a). The filter and the associated criteria of completeness imply an apparent modulation of the mean kinetic energy according to $\phi$, mainly for the peripheral collisions.
In the figure \ref{fig6}-a, the addition of the reconstruction effect involves a right-left skewness of the modulation, still related to the ``right-left effect''. This phenomenon is more important for alpha particles.
But it is also observed for the other light particles (p, d, t, $^3$He). It still exists when the studied physical quantity is not the mean kinetic energy but the temperature determined by maxwellian fits of the kinetic energy spectra.
\begin{figure} [htbp]
\centerline{\includegraphics[width=8.7cm, height=8.7cm] {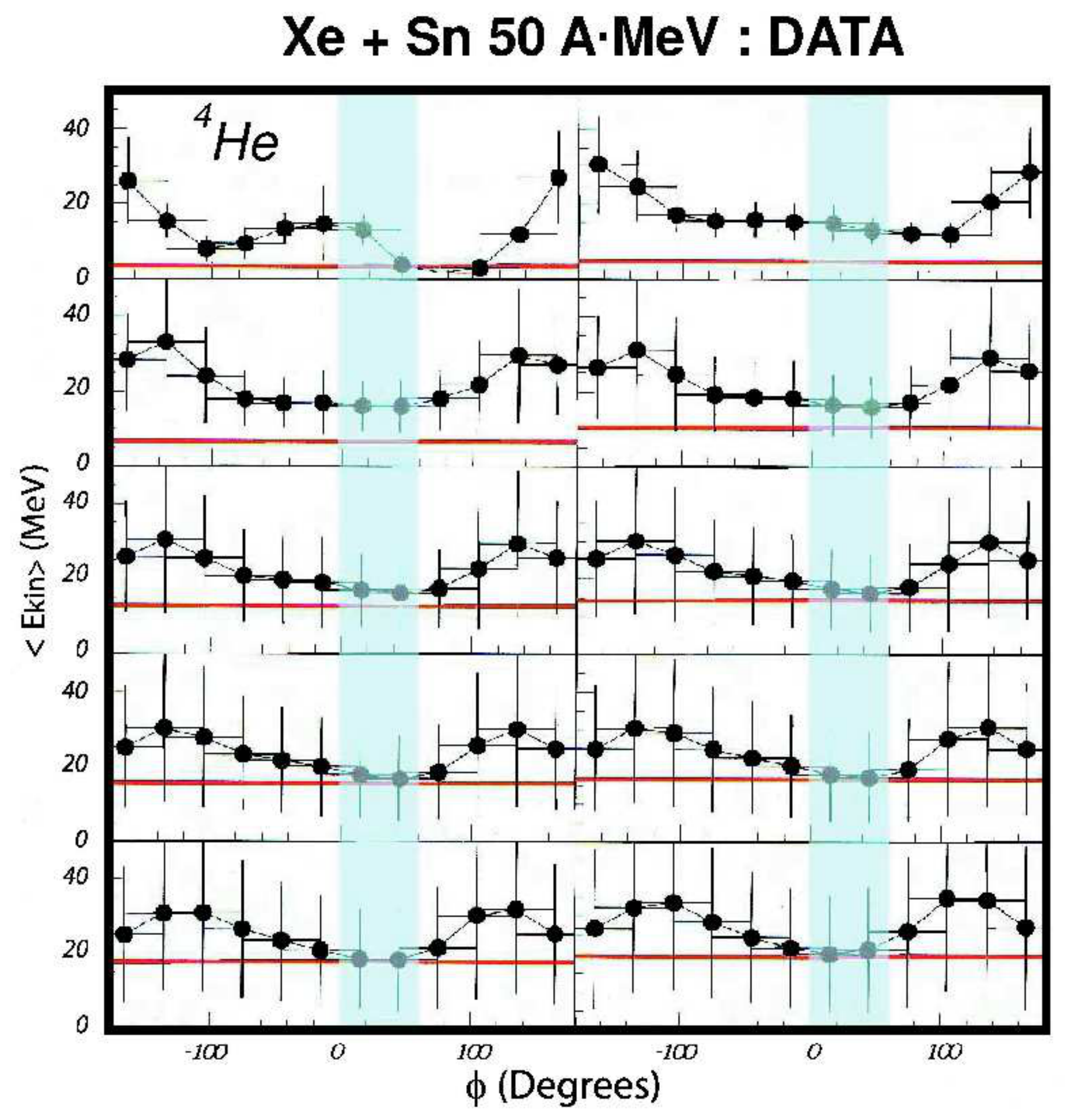}}
\caption{Mean energy distributions of the alphas evaporated by the QP according to $\phi$ and the various selections according to $E_{tr12}$ for the experimental data obtained with INDRA. The considered events are events known as ``complete''.  The red line corresponds to 2 times the measured temperature.\label{fig7}}
\end{figure}
Figure \ref{fig7} shows the experimental data to compare with the simulation results of the figure \ref{fig6}-a. There is an important difference  due to the pre-equilibrium emission around -180$ ^{\circ} $ / 180$ ^{\circ} $. Only the area 0$ ^{\circ} $- 60$ ^{\circ} $ in $\phi$, symbolized by a transparent light blue band, seems partly consistent with the mean kinetic energy expected because of evaporation as displayed by the horizontal red line. This trend is also found for the other light particles not shown here for the sake of simplicity. These observations are confirmed by figure \ref{fig8} where we show the experimental energetic spectra of tritons, defined in the reconstructed frame of the QP for various $\phi$ bins and a selection in transverse energy corresponding to semi-peripheral collisions. We see again that the particles emitted only in an azimuthal domain  between -30$ ^{\circ} $ and 60$ ^{\circ} $  give spectra compatible with a pure thermal emission. This is also seen for the other light particles.

Other studies have been made with the event generator HIPSE \cite{Lacroix1}, which treats more carefully the LCP productions at mid-rapidity than SIMON. They  also show that only a very limited angular domain in the front of the QP is almost not polluted \cite{Vient2}. This type of results had been also pointed out using Landau-Vlasov calculations but for smaller systems in the reference \cite{Eudes1}.
\begin{figure} [htbp]
\centerline{\includegraphics[width=8.9cm, height=8.9cm] {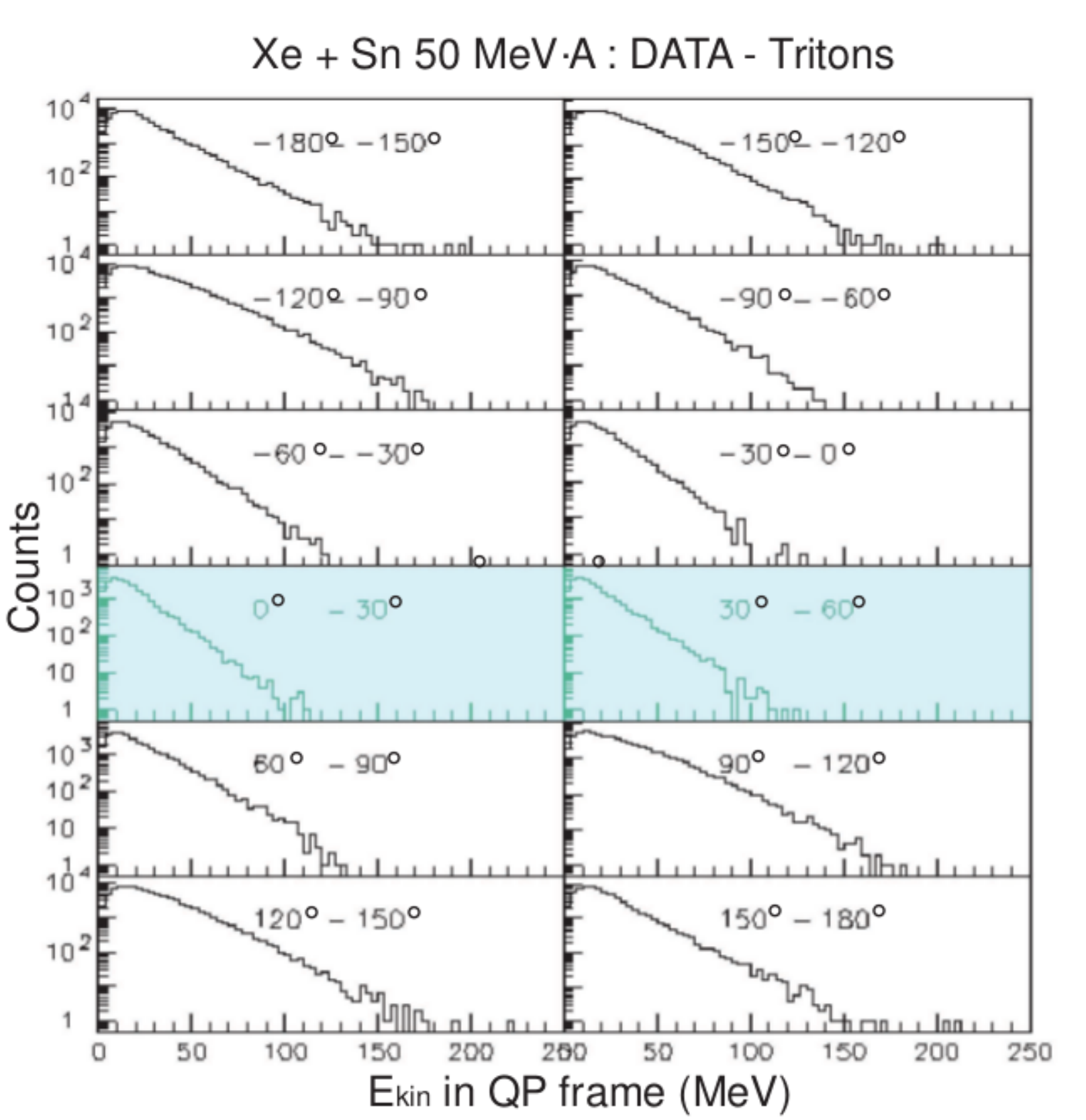}}
\caption{Kinetic energy distributions of the tritons in the reconstructed frame of the QP, for various angular selections according to $\phi$ obtained by INDRA collaboration for the system Xe + Sn at 50 A.MeV. The events are complete and correspond to semi-peripheral collisions.\label{fig8}}
\end{figure}
\subsection{The importance of selecting the reaction mechanism \label{ssec2.2}}
\begin{figure} [htbp]
 \centerline{\includegraphics[width=7.9cm, height=5.99cm] {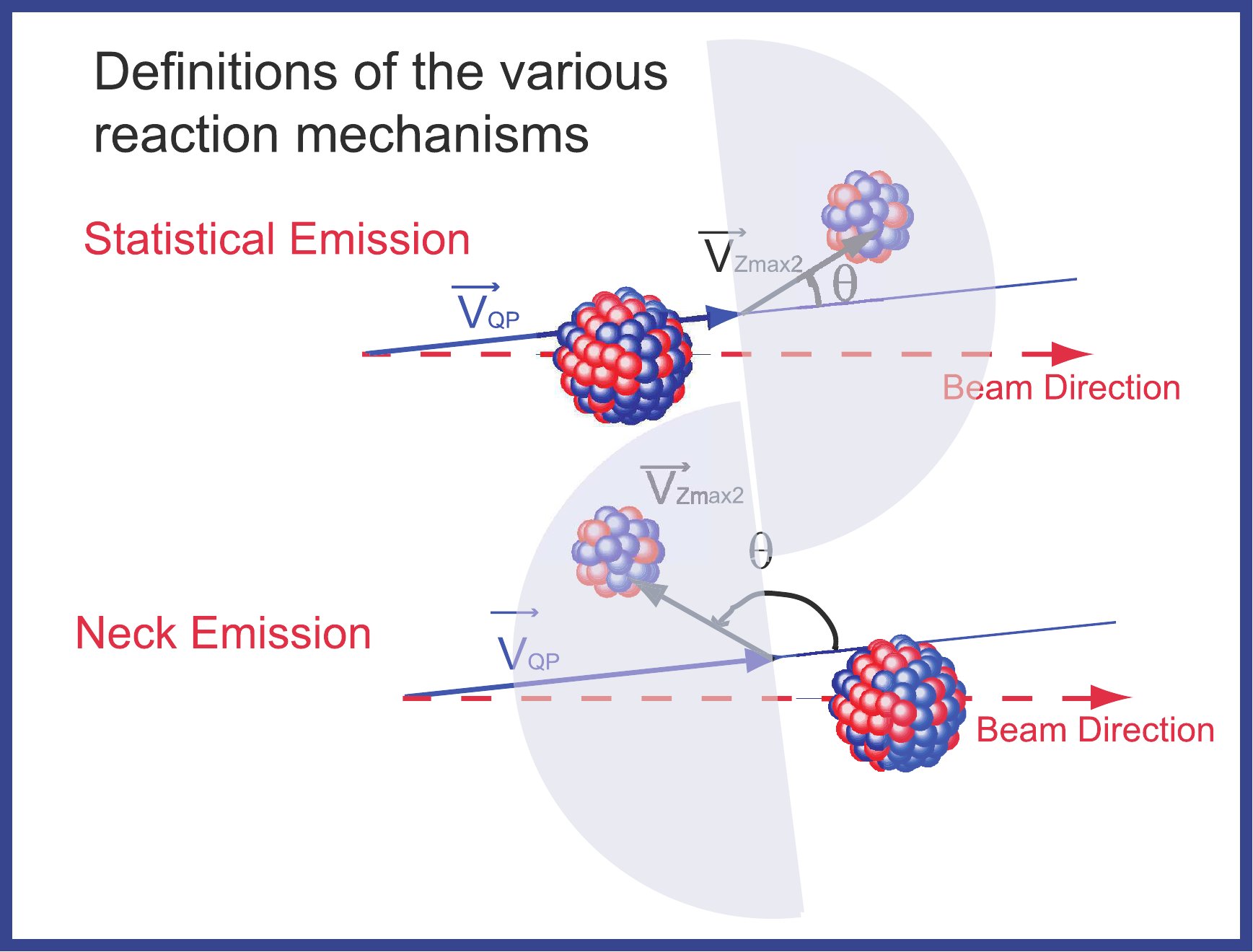}}
\caption{Diagram plotting the selection criteria of the two studied reaction mechanisms.\label{fig9}}
\end{figure}

\begin{figure} [htbp]
\centerline{\includegraphics[width=8.9cm, height=6.74cm] {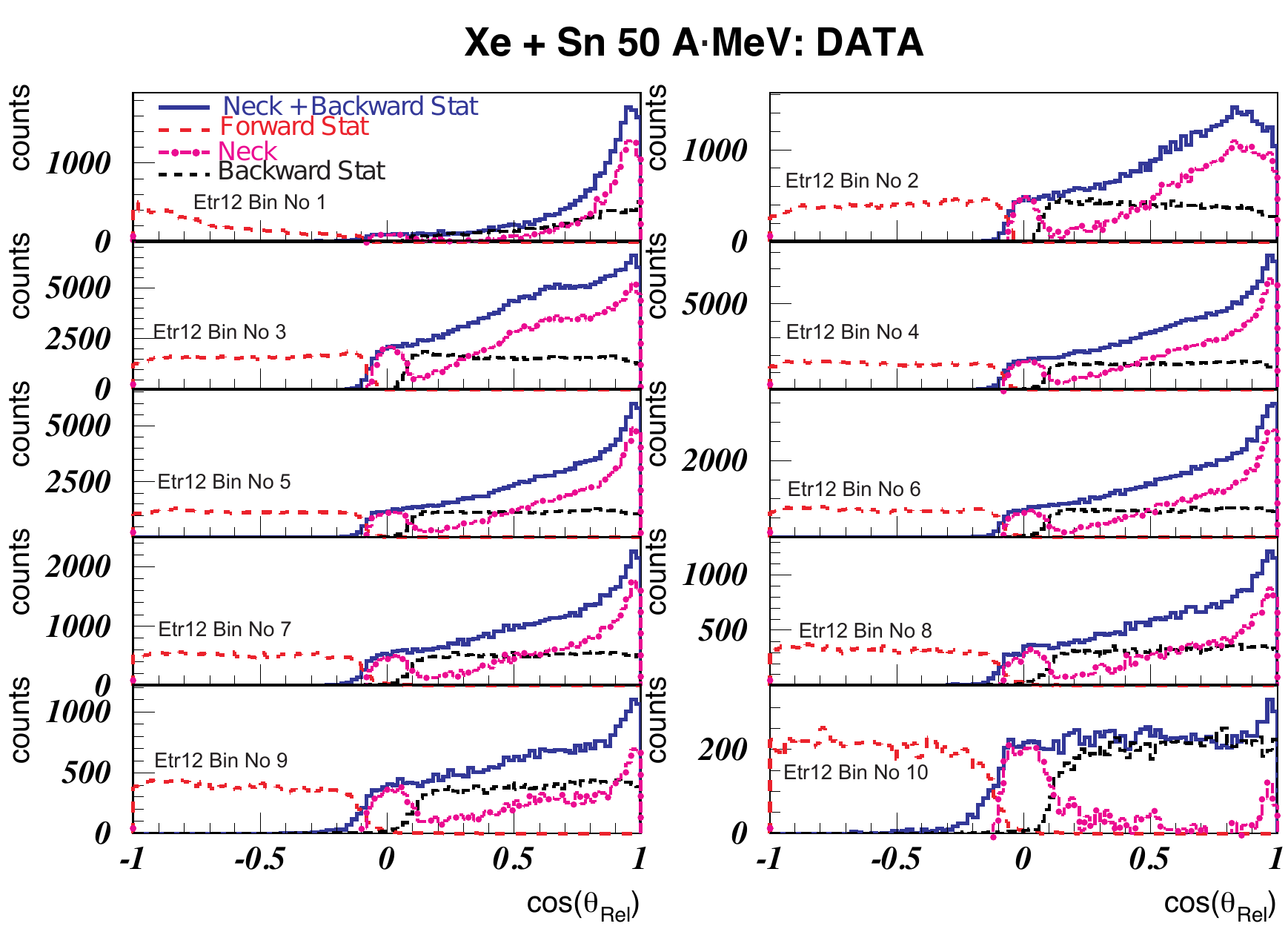}}
\caption{Distributions of the cosine of the angle $\theta_{rel}$, presented for events having a medium mass asymmetry between the two biggest fragments of the event and for various selections: the curve (blue full line) corresponds to the \textbf{Neck Emission}, the curve (red dashed line) at left corresponds to the \textbf{Statistical} emission in the front of the QP frame and this one (small black dashed line) at the right corresponds to the backward QP emission, the curve (pink line and point) corresponds to the \textbf{Pure Neck Emission} for which the statistical contribution was subtracted. The violence of the collision increases when one goes from left to right and downwards.\label{fig10}}
\end{figure}

As already discussed in the introduction, in heavy ion reactions at intermediate energy it is possible to observe different reaction mechanisms. We defined selection criteria in order to disentangle between pure binary collisions followed by a standard statistical decay and collisions with neck formation. 
 We will call the former \textbf{Statistical} collision and the latter \textbf{Neck Emission}. It was shown in \cite{Vient2, Colin1, Normand1} that a valid criterion, used to identify an event as \textbf{Statistical} collision, is  the presence of the second heaviest fragment in the front of the center of mass, in the forward hemisphere of the QP frame (see figure \ref{fig9}). \textbf{Neck Emission} events are those with mid-rapidity emission (nuclear matter between the two partners of the binary collision).  This discrimination allows to isolate  binary collisions followed by statistical emission. This latter seems to be a more adapted scenario with respect to the bases of our reconstruction method of the hot nucleus. 
 It was already shown and discussed in references \cite{Vient2, Colin1, Normand1} the fact that these two mechanisms correspond to well different dynamics. The process of energy dissipation may have different origins as reported in the reference \cite{Colin1}.
In  figure \ref{fig10}, we selected the violence of the collision using the variable $E_{tr12}$ normalized to the available energy in the center of mass in the direction perpendicular to the beam. We use also the variable $\eta$ which characterizes the charge asymmetry between the two heaviest fragments in the front of the center of mass.
\begin{equation}
\eta=(Z_{max1}-Z_{max2})/(Z_{max1}+Z_{max2})
\label{equa1}
\end{equation}
 This variable is used to obtain a signal of bimodality to characterize an eventual liquid-gas phase transition in nuclei \cite{Pichon1, Bonnet2, Bruno1}.  Figure \ref{fig10} shows, for various violences of collision (different $E_{tr12}$ domains), the distribution of the cosine $\theta _{rel}$, the angle formed between the relative velocity vector of the two heaviest fragments in the front of the center of mass  (see definition in equation \ref{equa2}) and the velocity vector of the reconstructed source. We limit ourselves to present a typical case here: collisions for the system Xe + Sn at 50 A.MeV, having a medium asymmetry between the two biggest fragments (\textit{i.e.}, $ 0.33 < \eta \leqslant 0.66$). We define furthermore:
 \begin{equation}
\overrightarrow{V_{rel}}=\overrightarrow{V}_{Z_{max1}}-\overrightarrow{V}_{Z_{max2}}
\label{equa2}
\end{equation}
We can note that the events, labeled as \textbf{Statistical}, present a flat distribution, if we neglect certain problems of angular acceptance, which are more visible for the peripheral collisions. The events of type \textbf{Neck} (blue full lines) present rather a focusing of the relative velocity vector in the direction of the velocity vector of the reconstructed source; this focusing effect is increasingly larger as the collision is less violent. To improve the selection of this contribution, we may subtract the statistical backward distribution, for which the second heaviest fragment would be emitted backward with respect to the QP. This back statistical distribution should be the symmetrical distribution of the events known as \textbf{Statistical}, such as we select them, with respect to the zero abscissa of the distribution (black dashed line).  It must give positive relative angle values taking into account the definition of the relative velocity vector.  We obtain thus after subtraction, a proper cosine $\theta_{rel} $ distribution of the events corresponding to the genuine \textbf{Neck Emission},\textit{i.e.} \textbf{Pure Neck Emission} (pink line and point in figure \ref{fig10}). The distribution shape does not change, but the proportion of events can be estimated more carefully; this contribution tends to disappear when the violence of the collision grows. In fact, we choose to keep both mechanisms to see if differences appear when we apply our new calorimetry.
\section{THE NEW ``3D CALORIMETRY''\label{sec3}}
\subsection{Determination of the spatial domain of emission by the QP\label{ssec3.1}}
\begin{figure}[ht]
\centerline{\includegraphics[width=8.3cm,height=5.03cm]{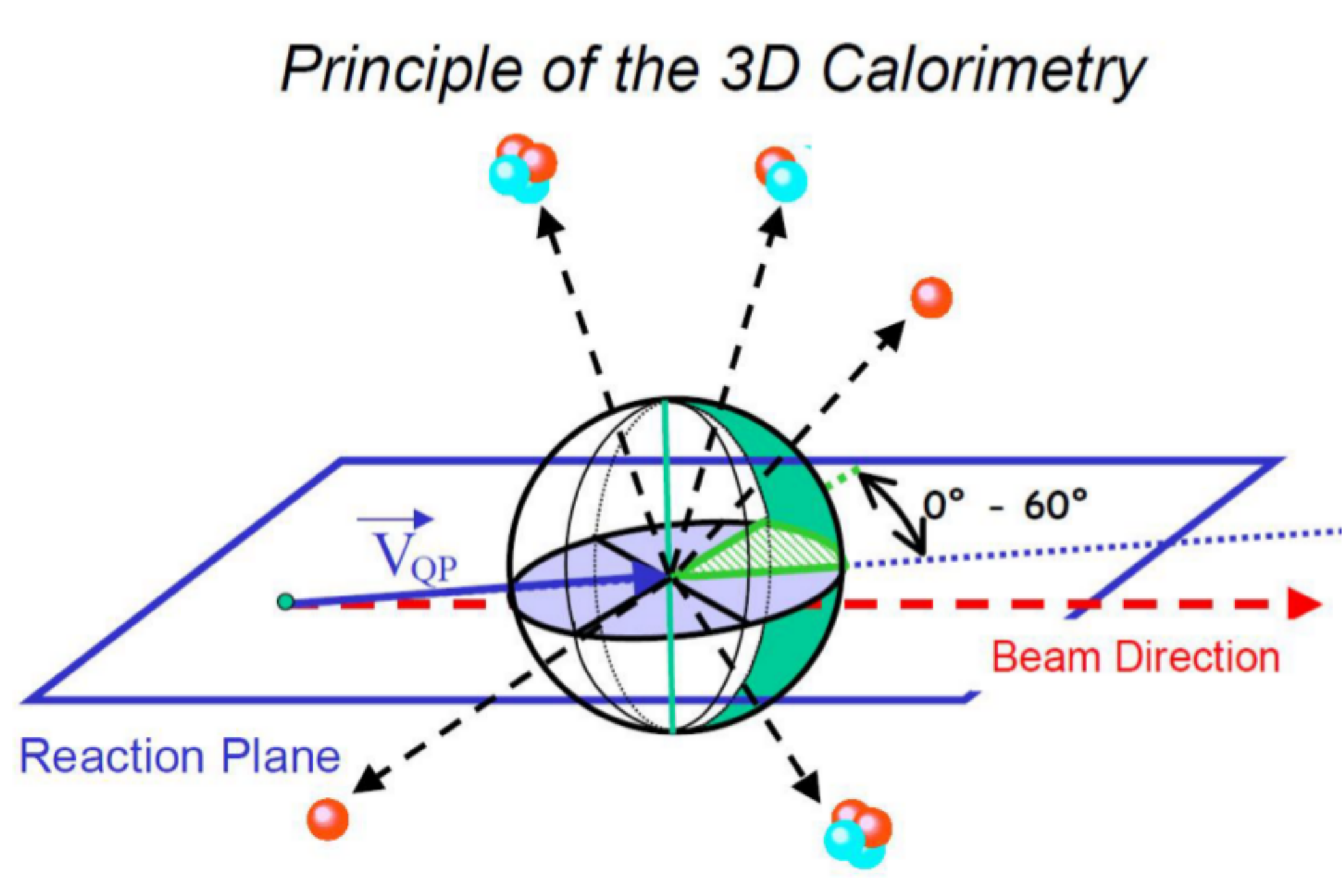}}
\caption{Diagram allowing to visualize the considered spatial domain to define the probability of emission by the QP for all types of particles.\label{fig11}}
\end{figure}

The study reported in the reference \cite{Vient1} and the conclusions drawn in the sub-section \ref{ssec2.1} convinced us that to correctly characterize the QP de-excitation, we must use a quite restricted spatial domain.
From the angular definitions of $\phi$ and of $\theta_{spin}$ given in the sub-section \ref{ssec2.1}, we consider \textbf { as particles emitted actually and solely by the QP: the particles located in the angular azimuthal domain included between 0$^{\circ}$ and 60$^{\circ}$ in the reconstructed QP frame}. It thus corresponds to one-sixth of the total solid angle as we can see it in figure \ref{fig11}. This spatial domain is, by definition, linked to the QP vector velocity reconstructed in the frame of the center of mass, and it will change according to the various selections on the violence of the collision. 
The Quasi-Projectile is then reconstructed, event by event, by assigning to each particle a given probability to be emitted, which was determined using all the information obtained in the spatial domain defined above. This kind of calorimetry was partially used in a simplistic way in \cite{Vient3}.
\subsection{Calculation of the emission probabilities by the QP\label{ssec3.2}}
 To apply this 3D calorimetry and define the emission probability by the QP, we assume that the process of de-excitation of the QP presents a symmetry of revolution around the perpendicular axis to the reaction plane described by the reconstructed velocity vector of the QP and the velocity vector of the initial projectile. We defined, for $\phi$ varying from -180$^{\circ}$ to 180$^{\circ}$, six areas 60$^{\circ}$ wide. At first, for each one, we built all the polar angular distributions as well as the energy distributions for all the types of detected particles, defined in the frame of the reconstructed QP.
  This is made by using all the light particles and the intermediate mass fragments  with the exception of the two heaviest fragments emitted in the front of the center of mass (the probability is one for both nuclei coming from the QP). From these distributions, for an angular domain between $\phi_{1} $ and $\phi_{2}$, we determine an experimental emission probability by the QP for a particle of kinetic energy $E_{k}$ at an angle $\theta_ {spin}$ and in an azimuthal angular domain  $\Delta\phi$: $Prob(E_{k},\theta_ {spin},\Delta\phi=\phi_{1}-\phi_{2})$.
  
 We assume at first that, for any particle, the probability to be emitted in a polar angle $\theta _ {spin} $ is independent from the kinetic energy and vice versa. This means that we neglect the influence of the angular momentum on the distributions of kinetic energy. Considering this hypothesis, we can thus calculate the probability from the following relation: 
\begin{equation}
Prob(E_{k},\theta _ {spin} ,\Delta\phi)=Prob(E_{k},\Delta\phi)\times Prob(\theta _ {spin},\Delta\phi)
\label{equa3}
\end{equation}
From this last relation and from our choice of selection of particles emitted by the QP, we deduce that the experimental probability is given  by the following relation:
\begin{equation}
Prob(E_{k},\theta _ {spin} ,\phi_{1}-\phi_{2})=\frac{\frac{dN(E_{k},0^{\circ}-60^{\circ})}{dE_{k}}}{\frac{dN(E_{k},\phi _{1}-\phi _{2})}{dE_{k}}} \times \frac{\frac{dN(\theta _ {spin} ,0^{\circ}-60^{\circ})}{d\cos \theta _ {spin} }}{\frac{dN(\theta ,\phi_{1}-\phi _{2})}{d\cos \theta _ {spin} }}
\label{equa4}
\end{equation}
\begin{figure}[htbp]\centerline{\includegraphics[width=8.5cm,height=9.5cm]{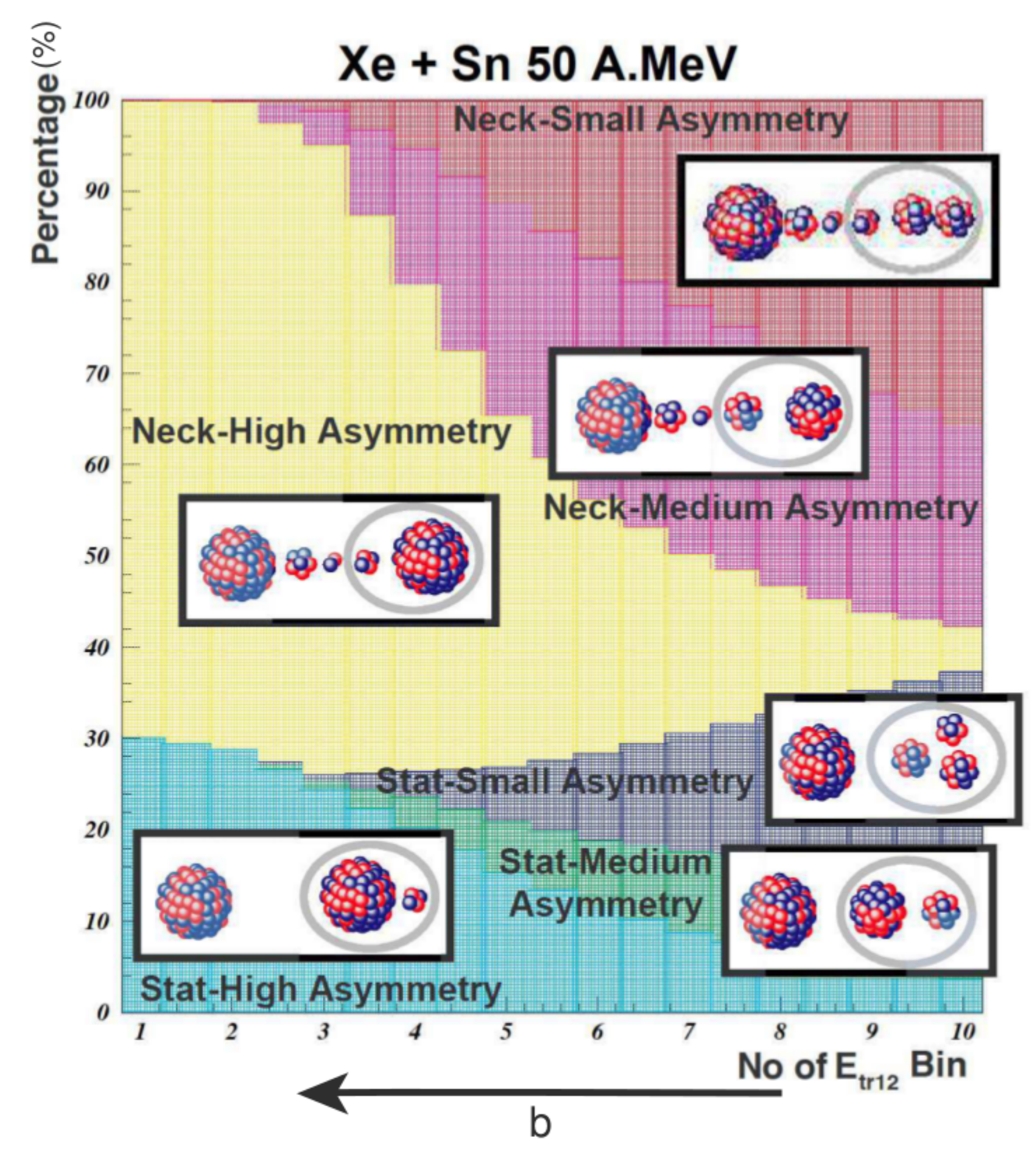}}
\caption {Proportions of the various types of reaction mechanisms according to the ``experimental impact parameter'' (violence of the collision) for collisions Xe + Sn at 50 A.MeV.\label{fig12}}
\end{figure} 
This calculation is done also according to the violence of the collision, the mechanism of reaction and the asymmetry between the two heaviest fragments in the front. Furthermore, the chosen completeness criteria are here less drastic than previously and focused on the reconstruction of the QP. The studied events are here complete events to the front of the center of mass. For these events one can observe in figure \ref{fig12}, the proportion of the different selections of mechanism and decay taken into account in this study on the basis of the violence of the collision. We observe a forward statistical contribution which is relatively constant around 30-35 $\% $ (we remind that this number must be doubled to get the right proportion of events corresponding to the statistical emission).

We thus obtain experimental probability functions depending on the following variables:

$Prob(E_{k},\theta_{spin},{\phi_{1}-\phi_{2}},E_{tr12} Norm,\eta, Mechanism,Z,A )$\\
for the particles of charge $ Z \leqslant 3$, identified also by mass.

$Prob(E_{k},\theta_{spin}, {\phi_{1}-\phi_{2}},E_{tr12} Norm,\eta,Mechanism,Z)$\\
for the particles of charge $Z>3$.

We can note that we assumed here that the probability of detection is independent of $E_{c}$, $\theta$ and $\phi$. 
\subsection{QP reconstruction\label{ssec3.3}}
For each particle ($ Z_{n}$, $ A_{n}$, $\overrightarrow{P_{n}}$) detected in an event, we determine its kinetic energy, the polar and azimuthal angle of its velocity vector, defined in the reconstructed frame. We deduce thus its probability $Prob_{n}$ to be emitted by the QP from the experimental functions of probability (see Eq \ref{equa4}). We then associate this probability to this particle to reconstruct the characteristics of the QP.
The QP charge can be estimated as follows:
\begin{equation}
Z_{QP}=\sum\limits_{n=1}^{multot}Prob_{n}\times Z_{n}
\label{equa5}
\end{equation}
For the mass, we need to make some hypotheses. We assume that the QP keeps the isotopic ratio of the initial projectile and that nuclei follow the valley of stability.
The mass conservation allows us to deduce the number of neutrons produced by the QP as indicated by the relation below.
\begin{equation}
A_{QP}=Z_{QP}\times 129/54=\sum\limits_{n=1}^{multot}Prob_{n}\times
A_{n}+N_{neutron}
\label{equa6}
\end{equation}
We can then deduce the reaction Q-Value.
\begin{equation}
\begin{split}
Q=E_{b}(A_{QP},Z_{QP})-\sum\limits_{n=1}^{multot}Prob_{n} \times E_{b}(A_{n},Z_{n})\\
-N_{neutron}\times E_{b}(1,0)
\end{split}
\label{equa7}
\end{equation}
$E_{b}(A_{n},Z_{n})$ is the binding energy of the nucleus $_{Z_{n}}^{A_{n}} X $.\\
We determine the QP velocity vector in the frame of the center of mass of the reaction only from charged particles by using the following expression:
\begin{equation}
\overrightarrow{V}_{QP}=\frac{\sum\limits_{n=1}^{multot}Prob_{n}\times 
\overrightarrow{P_{n}}}{(A_{QP}-N_{neutron})}
\label{equa8}
\end{equation}
with $\overrightarrow{P_{n}}$ the linear momentum of the $n^{th}$ particle in the frame of the c.m..\\
We can then calculate the QP excitation energy:
\begin{equation}
E_{QP}^{\ast}=\sum\limits_{n=1}^{multot}Prob_{n}\times
E_{kn}+N_{neutron}\times \left\langle E_{k}\right\rangle _{p+\alpha
}-Q-E_{kQP}
\label{equa9}
\end{equation}
being  $E_{kn}$ the kinetic energy of the $n^{th}$ particle in the frame of the center of mass, 
$\left\langle E_{k}\right\rangle_{p+\alpha}$ the mean kinetic energy of the neutrons deduced from those of the protons and alphas (Coulomb energy is subtracted) and finally with $E_{kQP}$ the  QP kinetic energy in the center of mass.
\section{Comparison with a classic calorimetry\label{sec4}}
 As a rough guide, we will compare this ``3D calorimetry'' with  the ``Standard Calorimetry'' based on a technique doubling the particle contribution in the front of the QP,  described in \cite{Viola2} \cite{Steck2} \cite{Vient1} \cite{Bonnet1}.

\begin{figure}[htbp]\centerline{\includegraphics[width=8.5cm,height=15.4cm]{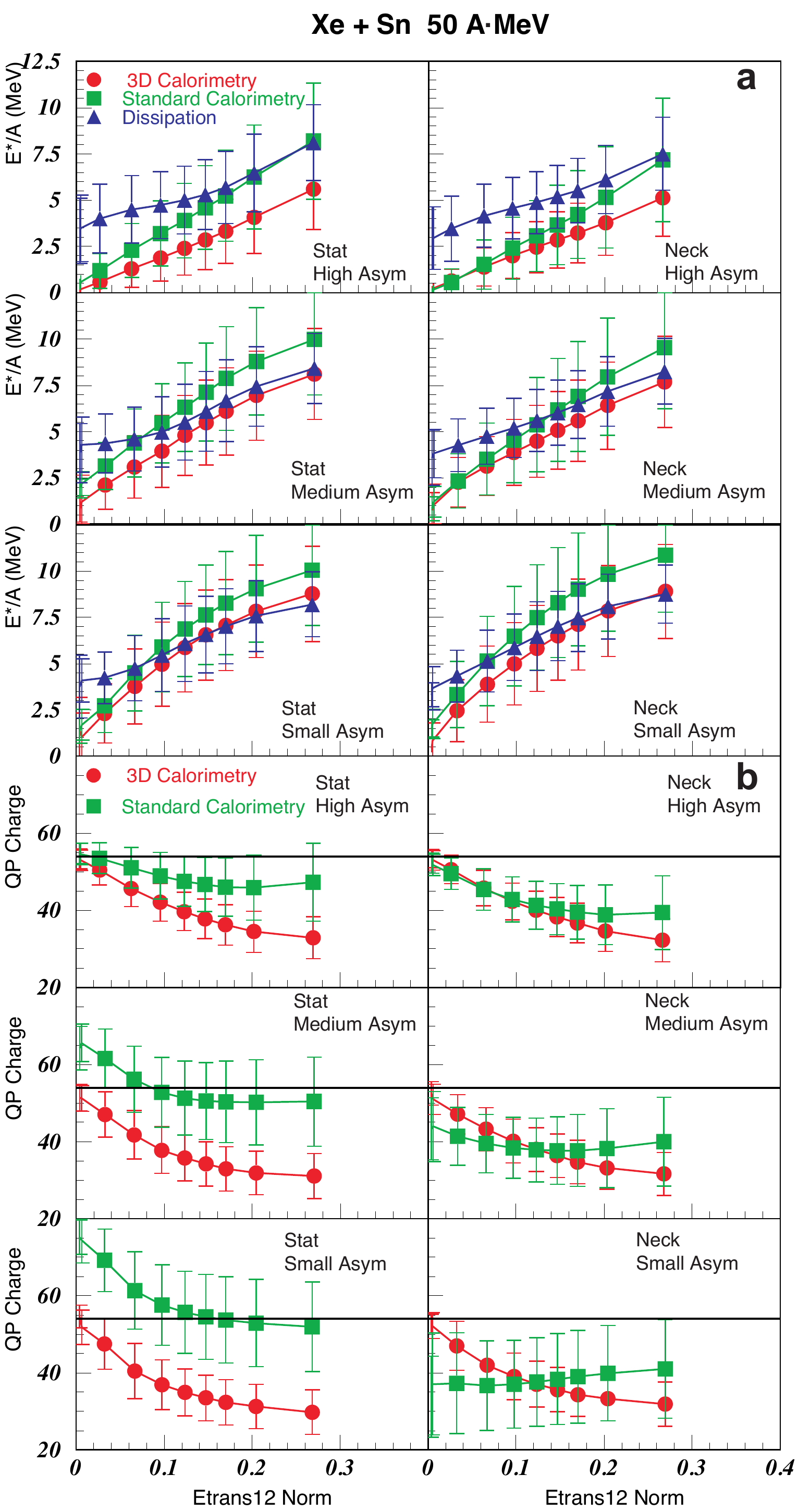}}
\caption {\textbf{a) }Average correlation between the measured QP excitation energy per nucleon and the normalized transverse energy of LCPs for various selections of mechanism and asymmetry for collisions Xe + Sn at 50 A.MeV.
\textbf{b)} Average correlation between the QP reconstructed charge and the normalized transverse energy of LCPs for various selections of mechanism and asymmetry for collisions Xe + Sn at 50 A.MeV (the black line indicates the projectile charge).\label{fig13}}
\end{figure}

  Figure \ref{fig13} shows the average evolution of the QP charge and excitation energy per nucleon for the various selections of interest.
Figure \ref{fig13}-a shows the excitation energy per nucleon for the two different calorimetries and the dissipated energy per nucleon during the collision. This latter  is in fact determined from the measured QP velocity by our method. As indicated in the reference \cite{Metivier1}, we can obtain the quantity of the initial incident energy in MeV, which is dissipated per nucleon during the collision by the following expression: 
\begin{equation}
\frac{E_{Dissipated}}{A}=\frac {E_{CM}}{(A_{Proj}+A_{Tar})}-\frac{1}{8} \times {\dfrac{V_{rel}^{2} }{c^{2} }}\times 1\,amu\,c^{2}
\label{equa10}
\end{equation}
with  $E_{CM}$ the available energy in the center of mass of the reaction and $V_{rel}$ the relative velocity between partners of the collision. 

Here, as the system is symmetric, we assume that $V_{rel} \simeq 2 \times V_{QP}$, with $V_{QP} $ velocity of the QP in the frame of the center of mass. The equation \ref{equa10} is valid only for symmetric systems and assumes that pre-equilibrium emission is symmetric in the frame of the center of mass. This energy thus represents the maximum energy which can be stored by the QP. Figure \ref{fig13}-a gives, for each kind of mechanism and each asymmetry, the curves relative to the excitation energy per nucleon obtained by the 3D calorimetry (red circles),  the standard one (green squares) and the dissipated energy per nucleon (blue triangles). In this figure,  one can notice immediately that there is a qualitative visible improvement of the measurements. With the new calorimetry, for the most peripheral collisions, we obtain lower limit values which seem reasonable: they are close to $E_{QP}^{\ast }=$ 0 A.MeV for the excitation energy per nucleon and $Z_{QP}=$ 54 for the charge of the QP  (figure \ref{fig13}-b), which is not always the case for the standard calorimetry. Moreover, contrary to the standard method the measured excitation energy never exceeds the estimated dissipated energy. There is thus a reasonable coherence between the measured QP velocity and the measured excitation energy per nucleon.

Regarding the measurement of the QP charge in Figure \ref{fig13}-b, the very different measurements, obtained with the standard calorimetry between the events called \textbf{Statistical} and the events with a \textbf{Neck Emission}, disappear completely with the new 3D calorimetry. The indicated standard deviations are also clearly smaller. 

This simple comparison is not naturally enough to validate this new calorimetry but gives some evidences that this 3D calorimetry improves a lot the measurement of QP ($E_{QP}^{\ast}$ and $ Z_{QP} $). It seems essential to study it by means of the most realistic possible simulation and this will be the subject of a next paper.
\section{CONCLUSIONS\label{sec5}}
By provoking extremely violent collisions of heavy ions, nuclear physicists try to modify the internal energy of nucleus.  In the domain of Fermi energy how this energy is deposited and stored is still a subject of many discussions. This is mainly due to the fact that it is very difficult to prove experimentally in an unquestionable way that a hot nucleus, thermodynamically well equilibrated, was formed. In this context, we described the foundations of a method of QP reconstruction and reminded difficulties which can intervene.

We chose to follow an experimental approach trying to solve us gradually the difficulties. For that purpose, we used the event generator SIMON and a simulation program, reproducing as accurately as possible the behavior of our experimental device.

First, being placed in the framework of binary collisions generated by SIMON without pre-equilibrium, we study the de-excitation of an isolated hot Quasi-Projectile. We observed important spatial and energetic distortions concerning the evaporated particles by the QP. We showed the fundamental roles played by the experimental device and the recoil effects. 

To make a correct calorimetry, we need to detect all the evaporated fragments and particles by the QP. Therefore, we must use events said ``complete events''.  For the very peripheral collisions, the detection of the QP residue remains difficult because of the INDRA forward acceptance. For collisions when few particles are evaporated, they must have a large transverse linear momentum to deviate correctly the QP residue allowing its detection. ``Complete events'' correspond mainly to these type of event for peripheral collisions. It is difficult to compensate for the recoil effect due to the emission of the first evaporated particle. Consequently, we do not find isotropic spatial and energetic distributions but a recoil effect called  ``right-left'', which is dominant for peripheral collisions. It involves indirectly also difficulties in the determination of the source velocity, when only IMFs and heavy fragments are used to reconstruct the frame of the hot nucleus. There is clearly a difference between the center of mass of the IMFs and that of the LCPs, which involves an apparent energetic contribution of LCPs too large in the reconstructed frame of the QP. All these facts are confirmed with the real experimental data and even amplified by pre-equilibrium contribution.
  
Second, we presented a new calorimetric protocol to characterize the Quasi-Projectile. This calorimetry is based on the experimental determination of an emission probability starting from the physical characteristics of particles in a restricted domain of the velocity space. We wanted that this one can take into account the influence of  pre-equilibrium particles and of a possible contribution of the mid-rapidity such the Neck Emission. 

In this study, we added supplementary criteria of event selections to observe the robustness of this calorimetry. We differentiated the collisions with Neck Emission from the others. We also took into account the asymmetry between the two biggest fragments in the front of the center of mass. 

 To demonstrate completely the interest of the 3D calorimetry, we compared it with a standard method consisting doubling of the light charged particles located in the front of the reconstructed frame. This last one seems clearly inaccurate. They tend to give excitation energies per nucleon too large in comparison to the apparent dissipation, as it can be observed clearly in the figure \ref{fig13}. They also provide  too wide distributions of the excitation energy per nucleon or the charge for the reconstructed QP \cite{Vient1}. This brings moreover into question the quality of selections done with an excitation energy obtained with standard calorimetric methods like this. The new calorimetry gives better estimates of the excitation energy per nucleon than previous methods. But it can allow, by its intrinsic hypotheses (two particle sources), to make solely a correct physical characterization of binary collisions, \textit{i.e.} peripheral or semi-peripheral collisions.

To complete our conclusion on this study, we can notice that a real improvement of the quantitative characterization of the QP can be obtained only, either by an effective correction of the complex distortions generated by the experimental device and the criteria of event selection, or by the construction of a 4$\pi$ detector, which would have a higher granularity, a forward efficient detection and an excellent isotopic resolution, improving the identification and the kinematic characterization of particles. We finally have to keep in mind that all these conclusions can be drawn  in principle only for the studied system, Xe + Sn, and are to be confirmed for the others. This work confirms on the other hand in a clear way that a comparison between a theoretical model and experimental data makes sense in this domain of physics only if the theoretical model is passed through a software filter simulating the totality of the detector response.

\newpage 
\bibliography{calorimetryPartI}
\end{document}